\documentclass[aps,pra,reprint,nofootinbib,twocolumn,superscriptaddress,showpacs,showkeys,longbibliography]{revtex4-1}

\usepackage{eurosym}
\usepackage{amsmath,amssymb,amstext}
\usepackage{ulem}
\usepackage[usenames,dvipsnames]{color}
\usepackage{graphicx}
\usepackage{braket}
\usepackage{natbib}
\usepackage{comment}
\usepackage{dcolumn}
\usepackage[english]{babel}
\usepackage{wasysym}
\usepackage{bm}
\usepackage[colorlinks,bookmarks=false,citecolor=blue,linkcolor=red,urlcolor=blue]{hyperref}

\usepackage{appendix}
\usepackage{CJK}

\begin{document}
\title{Enantio-conversion of chiral mixtures via optical pumping}
\author{Chong Ye}\email{yechong@bit.edu.cn}
\affiliation{Beijing Key Laboratory of Nanophotonics and Ultrafine Optoelectronic Systems, School of Physics, Beijing Institute of Technology, 100081 Beijing, China}
\affiliation{Beijing Computational Science Research Center, Beijing 100193, China}
\author{Bo Liu}
\affiliation{Beijing Computational Science Research Center, Beijing 100193, China}
\author{Yu-Yuan Chen}
\affiliation{Beijing Computational Science Research Center, Beijing 100193, China}
\author{Yong Li}\email{liyong@csrc.ac.cn}
\affiliation{Beijing Computational Science Research Center, Beijing 100193, China}
\affiliation{Synergetic Innovation Center for Quantum Effects and Applications, Hunan Normal University, Changsha 410081, China}

\begin{abstract}
  Enantio-conversion with the help of electromagnetic fields is an essential issue due to the chirality-dependence of many chemical, biological, and pharmaceutical processes.
  Here, we propose a method for this issue based on a five-level double-$\Delta$ model of chiral molecules.
  By utilizing the breaking of left-right symmetry in the two $\Delta$-type sub-structures, we can establish the chiral-state-selective excitations with one chiral ground state being excited to an achiral excited state and the other one being undisturbed. In the meanwhile, the achiral excited state will relax to the two chiral ground states. The two effects simultaneously acting on the chiral mixtures can convert
  molecules of different chiralities to the ones of the same chirality, i.e., the enantio-conversion via optical pumping. We numerically show that highly efficient enantio-conversion can be achieved. Our method works in the appearance of decoherences and without the
  precise control of pulse-durations (pulse-areas) {or} pulse-shapes. These advantages offer it promising features in promoting the future exploring of enantio-conversion.

\end{abstract}
\date{\today}
\maketitle
\section{Introduction}
Recently, the inner-state enantio-purification~\cite{PRL.87.183002,PRA.77.015403,JPB.43.185402,PRL.118.123002,
Angew.Chem.56.12512,PRL.122.173202,PRA.100.043413,
PRAp.13.044021,PRA.101.063401,arxiv1,JCP.151.014302,PRA.100.043403,PRL.90.033001,PRL.84.1669,
PRA.65.015401,JCP.115.5349,JPB.37.2811,PRR.2.033064,arXiv2}, spatial enantio-separation~\cite{DA,DA1,PRL.99.130403,JCP.132.194315,JCP.137.044313,PRL.100.213004,PRL.93.032508,PRL.118.193401,
PRL.121.173002,YAO,PRL.122.024301}, and enantio-discrimination~\cite{PRA.84.053849,PJASB.88.120,Nature.497.475,PRL.111.023008,
PCCP.16.11114,ACI,JCP.142.214201,JPCL.6.196,JPCL.7.341,
PRA.100.033411,JCP.152.204305,Lehmannbook,Lehmann1,Lehmann3,PRL.117.033001,XuXW,OL.43.000435,PRA.99.023837,PRL.122.103201} have become essential issues, since the vast majority of chemical~\cite{A1}, biological~\cite{A2,A3,A4}, and pharmaceutical~\cite{A5,A7,A8,A6} processes essentially depend on molecular chirality.
For these purposes, the (electronic, vibrational, {or} rotational) inner states of chiral molecules are manipulated with the help of electromagnetic fields.
Moreover, precisely manipulating the rotational populations of chiral molecules has opened new avenues to study parity violation~\cite{A9,A90}.

The inner-state enantio-purification includes
the enantio-specific state transfer~\cite{PRL.87.183002,PRA.77.015403,JPB.43.185402,PRL.118.123002,
Angew.Chem.56.12512,PRL.122.173202,PRA.100.043413,
PRAp.13.044021,PRA.101.063401,arxiv1,JCP.151.014302,PRA.100.043403} and the enantio-conversion~\cite{PRL.90.033001,PRL.84.1669,PRA.65.015401,JCP.115.5349,JPB.37.2811,PRR.2.033064,arXiv2}.
It %\cite{PRL.87.183002,PRA.77.015403,JPB.43.185402,PRL.118.123002,
%Angew.Chem.56.12512,PRL.122.173202,PRA.100.043413,
%PRAp.13.044021,PRA.101.063401,arxiv1,JCP.151.014302,PRA.100.043403,
%PRL.90.033001,PRL.84.1669,PRA.65.015401,JCP.115.5349,JPB.37.2811,PRR.2.033064,arXiv2}
aims to enhance the population excess of one chirality over the other
in an inner state in chiral mixtures. The achieved inner-state enantiomeric excess
characterizes the efficiency of the inner-state enantio-purification.
The enantio-specific state transfer~\cite{PRL.87.183002,PRA.77.015403,JPB.43.185402,PRL.118.123002,
Angew.Chem.56.12512,PRL.122.173202,PRA.100.043413,
PRAp.13.044021,PRA.101.063401,arxiv1,JCP.151.014302,PRA.100.043403} achieves the enhancement of inner-state enantiomeric excess without changing the chiralities of molecules.
{The enantio-conversion~\cite{PRL.90.033001,PRL.84.1669,PRA.65.015401,JCP.115.5349,JPB.37.2811,PRR.2.033064,arXiv2} is more ambitious, since it aims to convert molecules of different chiralities to the ones of the same chirality.} %\textcolor{blue}{After inner-state enantio-purification, the enantio-pure molecules} can be separated from the chiral mixtures by using energy-dependent processes (e.g. resonantly enhanced multi-photon ionization~\cite{PRL.122.173202}).

Most proposals of the inner-state enantio-purification~\cite{PRL.87.183002,PRA.77.015403,JPB.43.185402,PRL.118.123002,
Angew.Chem.56.12512,PRL.122.173202,PRA.100.043413,
PRAp.13.044021,PRA.101.063401,arxiv1,JCP.151.014302,PRA.100.043403,PRL.90.033001,
PRL.84.1669,PRA.65.015401,JCP.115.5349,JPB.37.2811,PRR.2.033064,arXiv2} are based
on few-level models  with left-right symmetry-breaking $\Delta$-type (sub-)~structures (e.g. the three-level $\Delta$ type model~\cite{PRL.87.183002,PRA.77.015403,JPB.43.185402,PRL.118.123002,
Angew.Chem.56.12512,PRL.122.173202,PRA.100.043413,
PRAp.13.044021,PRA.101.063401,arxiv1,JCP.151.014302,PRA.100.043403,PRL.90.033001} and the four-level double-$\Delta$ model~\cite{PRL.84.1669,PRA.65.015401,JCP.115.5349,JPB.37.2811,PRR.2.033064,arXiv2}).
In these models, the chirality-dependency
results from the sign difference between the products of the three electric-dipole transition moments in the $\Delta$-type (sub-)~structures related to
different chiralities. These models use
only electric-dipole interactions.
This property offers them advantages over the traditional chirality-dependent models usually used in enantio-discrimination~\cite{ME1,ME2,ME3}, since the usually weak
magnetic-dipole interactions were included in
the traditional ones~\cite{ME1,ME2,ME3}. Therefore, these models are used not only in
the enantio-discrimination~\cite{PRA.84.053849,PJASB.88.120,Nature.497.475,PRL.111.023008,
PCCP.16.11114,ACI,JCP.142.214201,JPCL.6.196,JPCL.7.341,
PRA.100.033411,JCP.152.204305,Lehmannbook,Lehmann1,Lehmann3,PRL.117.033001,XuXW}, but also
in
the inner-state enantio-purification~\cite{PRL.87.183002,PRA.77.015403,JPB.43.185402,PRL.118.123002,
Angew.Chem.56.12512,PRL.122.173202,PRA.100.043413,
PRAp.13.044021,PRA.101.063401,arxiv1,JCP.151.014302,PRA.100.043403,PRL.90.033001,
PRL.84.1669,PRA.65.015401,JCP.115.5349,JPB.37.2811,PRR.2.033064,arXiv2} and the spatial enantio-separation~\cite{PRL.99.130403,JCP.132.194315}.

Theoretical works of the inner-state enantio-purification~\cite{PRL.87.183002,PRL.90.033001,PRA.77.015403,JPB.43.185402,PRA.101.063401,
JCP.151.014302,PRA.100.043413,PRL.122.173202,PRA.100.043403,PRR.2.033064,arXiv2,arxiv1,PRAp.13.044021} based
on these models focused on designing the relative phases {or} intensities among the electromagnetic fields for perfect (or highly efficient) inner-state enantio-purification. This purpose was achieved by using adiabatical method~\cite{PRL.87.183002,PRL.90.033001}, separate-pulse method~\cite{PRA.77.015403,JPB.43.185402,PRA.101.063401,JCP.151.014302,PRA.100.043413}, shortcuts-to-adiabaticity method~\cite{PRL.122.173202}, dark-state method~\cite{PRA.100.043403,PRR.2.033064,arXiv2}, and
Raman-pulse method~\cite{arxiv1,PRAp.13.044021}.
Recently, based on the three-level $\Delta$-type model of chiral molecules,
there were breakthrough experiments for achieving enantio-specific state transfer~\cite{PRL.118.123002,Angew.Chem.56.12512} and enantio-discrimination~\cite{Nature.497.475,PRL.111.023008,PCCP.16.11114,
ACI,JCP.142.214201,JPCL.6.196,JPCL.7.341} in gas-phase samples by manipulating
the rotational states of chiral molecules with the help of electromagnetic fields.

In the experiments of enantio-specific
state transfer~\cite{PRL.118.123002,Angew.Chem.56.12512}, only small enhancement of enantiomeric excess has been achieved.
This is due to limitation factors such as the phase mismatching~\cite{Lehmann3}, the magnetic degeneracy~\cite{Lehmann3,JCP.137.044313}, and the thermal distributions of the rotational states~\cite{Lehmann3}.
The methods for solving the problems of the magnetic degeneracy~\cite{PRA.98.043403,JCP.151.014302} and the thermal distributions~\cite{Zhang} have been theoretically discussed.
{In addition,} the decoherences as well as
the requirement of the precise control of the pulse-durations (pulse-areas)~\cite{PRA.77.015403,JPB.43.185402,PRA.101.063401,JCP.151.014302,PRA.100.043403,PRA.100.043413} {or} pulse-shapes~\cite{PRL.87.183002,PRL.90.033001,PRL.122.173202} play as important limitations in achieving perfect (or highly efficient) inner-state enantio-purification.

In order to solve the problems related to the decoherences and the precise control of the pulses,
we look for the possibility of using optical pumping for enantio-conversion in this paper.
Optical pumping~\cite{RMP} is a widely used method for manipulating the inner states of
quantum targets, ranging from the nuclear spin of atoms~\cite{RMP1},
quantum dot~\cite{PRL.91.017402},
hyperfine states of atoms~\cite{PRL.122.203202} and molecules~\cite{PRL.98.133001},
to ro-vibrational (or rotational) states of molecules~\cite{PRL.109.183001,arXiv:2004.02848,PRX.10.031022}.
Excluding the details of different quantum targets, the key point of optical pumping is the state-selective
excitations~\cite{RMP}. The undesired state can jump to an excited state, and the desired state cannot. This is usually attributed to the differences in the transition frequencies {or} transition dipoles corresponding to the undesired-excited and desired-excited transitions~\cite{RMP}.
Simultaneously, the excited state(s) relaxes to the desired and undesired states~\cite{RMP}. Their combining effect will eventually convert the undesired state to the desired one~\cite{RMP}.
In this sense, the relaxations play a positive role in manipulating the inner states and
it is not necessary to precisely control the pulses
in the optical pumping.

For the sake of enantio-conversion via optical pumping, the chiral-state-selective excitations are needed. Specifically, the undesired chiral state is expected to {be} excited to an achiral excited state(s), and in the meanwhile the desired chiral state (the inversion image of the undesired one) is expected to {be} undisturbed. The chiral-state-selective excitations and the relaxations
act simultaneously on the chiral mixture and can eventually evoke the enantio-conversion via
optical pumping: most populations will stay in the desired chiral state finally.
{At first glance}, the chiral-state-selective excitations seem to {be
impossible, because} a chiral state and its inversion image are
{degenerate} by neglecting the tiny energy difference due to the fundamental weak force~\cite{A9}.
Further, the electric transition dipoles related to different chiralities only have
possible sign-differences~\cite{PRA.98.043403,JCP.151.014302,PRL.87.183002,JCP.1190.5105}.

We propose to realize the chiral-state-selective excitations based on a five-level double-$\Delta$ model
of chiral molecules (see Fig.~\ref{Fig1}). By well designing the relative phases and intensities among the electromagnetic fields, the desired chiral ground state is an eigenstates of
the system and thus is undisturbed. The undesired one is usually not an eigenstates of
the system due to the breaking of left-right symmetry in the $\Delta$-type sub-structures.
It can be excited to the achiral excited state. Specifically, we use the adiabatical-elimination technology~\cite{PRL.99.130403} to design the electromagnetic fields.
Based on this, we numerically show that highly efficient enantio-conversion can be achieved via optical pumping.

\section{Five-level double-$\Delta$ model}\label{MODEL}
\begin{figure}[h]
  \centering
  % Requires \usepackage{graphicx}
  \includegraphics[width=0.9\columnwidth]{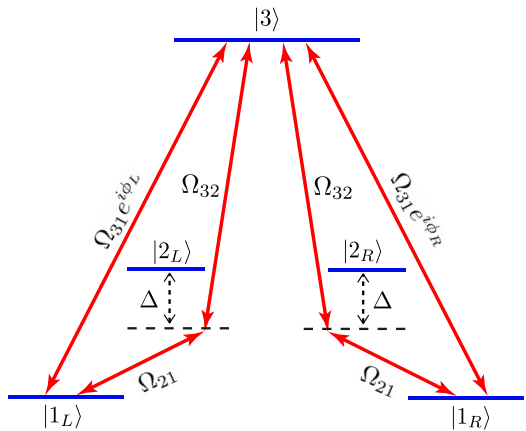}\\
  \caption{ The five-level double-$\Delta$ model of chiral molecules. The chiral ground states $|1_{Q}\rangle$, the chiral mediate-energy states $|2_{Q}\rangle$ and the achiral excited state $|3\rangle$ are coupled with three electromagnetic fields in the $\Delta$-type sub-structures with electric-dipole transitions $|1_{Q}\rangle\leftrightarrow|2_{Q}\rangle\leftrightarrow|3\rangle\leftrightarrow|1_{Q}\rangle$ in three-photon resonance condition. Here $Q~(=L,R)$ denotes the chirality. The corresponding coupling strengths are $\Omega_{21}$, $\Omega_{32}$, and $\Omega_{31}e^{i\phi_{Q}}$ with $\Omega_{ij}>0$. The chirality of the two $\Delta$-type electric-dipole sub-structures is reflected in $\phi_{L}=\phi$ and $\phi_{R}=\phi+\pi$.
    }\label{Fig1}
\end{figure}

We consider the five-level double-$\Delta$ model of chiral molecules as shown in Fig.~\ref{Fig1}. Our working states are two {degenerate} (left-handed and right-handed) chiral ground states $|1_L\rangle$ and $|1_R\rangle$, two {degenerate} chiral mediate-energy states $|2_{L}\rangle$ and $|2_{R}\rangle$, and one achiral excited state $|3\rangle$. Here, the subscript $Q~(=L,R)$ denotes the chirality.
The energies of the working states $|1_{Q}\rangle$, $|2_{Q}\rangle$, and $|3\rangle$ are $\hbar \omega_1$, $\hbar \omega_2$, and $\hbar \omega_3$, respectively. They are coupled via electric-dipole transitions and form two $\Delta$-type sub-structures $|1_{Q}\rangle\leftrightarrow|2_{Q}\rangle\leftrightarrow|3\rangle\leftrightarrow|1_{Q}\rangle$.
The frequencies of the three  electromagnetic fields are  $v_{21}$, $v_{32}$, and $v_{31}$, respectively.

%$\Omega_{32}$, $\Omega_{31}e^{i\phi_{L}}$, and $\Omega_{21}$, $\Omega_{31}e^{i\phi_{R}}$

We are interested in {the case of the} three-photon resonance. For simplicity, we assume the one-photon resonance of $|1_{Q}\rangle\leftrightarrow|3\rangle$. These
result in
\begin{align}
\Delta\equiv\Delta_{21}=-\Delta_{32},~~\Delta_{31}=0.
\end{align}
The detunings are $\Delta_{ij}\equiv \omega_{i}-\omega_{j}-v_{ij}$ with $i,j=1,2,3$ and $i> j$.
By using the rotating-wave approximation, the five-level double-$\Delta$ model can be described
by the Hamiltonian in the interaction picture ($\hbar=1$)
\begin{align}\label{5HT}
\hat{H}=&\sum_{Q}[\Delta|2_{Q}\rangle\langle 2_{Q}|+(\Omega_{21}|1_{Q}\rangle\langle 2_{Q}|+\Omega_{32}|2_{Q}\rangle\langle 3|\nonumber\\
&+\Omega_{31}e^{i\phi_{Q}}|1_{Q}\rangle\langle 3|+\mathrm{H.c.})].
\end{align}
The chirality is reflected in the overall phases~\cite{PRL.84.1669,JCP.115.5349,JPB.37.2811}
\begin{align}\label{PHID}
\phi_{L}\equiv\phi,~~~~\phi_{R}=\phi+\pi.
\end{align}
For the sake of simplicity and without loss of generality, we have assumed that
all $\Omega_{ij}>0$.

In the following, we will discuss the physical realization of the five-level double-$\Delta$ model.
In the gas-phase experiments of enantio-discrimination~\cite{Nature.497.475,PRL.111.023008,PCCP.16.11114,ACI,JCP.142.214201,
JPCL.6.196,JPCL.7.341} and the enantio-specific state transfer~\cite{PRL.118.123002,Angew.Chem.56.12512},
the rotational sub-levels are introduced~\cite{PRA.98.043403,JCP.151.014302,JCP.137.044313,
Nature.497.475,PRL.111.023008,PCCP.16.11114,ACI,JCP.142.214201,
JPCL.6.196,JPCL.7.341,PRL.118.123002,Angew.Chem.56.12512} in order to consider the molecular rotations.
We focus on the case of {chiral asymmetric-top molecules}.
For these chiral molecules, when the rotational sub-levels of the working states as well
as the polarizations and the frequencies
of the three electromagnetic fields are well designed, the involvement of rotational sub-levels will not
challenge the few-level models~\cite{PRA.98.043403,JCP.151.014302}.
Specifically, we can choose the rotational sub-levels of $|1_{Q}\rangle$, $|2_{Q}\rangle$, $|3\rangle$ to be
$|J_{\mathrm{k_a k_c M}}=0_{000}\rangle$,
$|1_{010}\rangle$, and
$(|1_{101}\rangle+|1_{10-1}\rangle)/\sqrt{2}$~\cite{PRA.98.043403,JCP.151.014302}, respectively. $|J_{\mathrm{k_a k_c M}}\rangle$ are the eigenstates of {asymmetric-top} molecules~\cite{PRA.98.043403,JCP.151.014302}.
The three electromagnetic fields coupling with the electric-dipole transitions $|1_{Q}\rangle\leftrightarrow|2_{Q}\rangle\leftrightarrow|3\rangle\leftrightarrow|1_{Q}\rangle$
are $Z$-, $X$-, and $Y$-polarized in the space-fixed frame, respectively.

Here, we will specify the vibrational sub-levels of the working states.
The chirality of molecules is related to the double-well--type energy surface potential in a vibrational degree of freedom.
One chiral state and its {degenerate} inversion image are localized in the two wells, respectively. {They are not the eigenstates of the double-well potential. The tunneling
between them would make the enantio-purified molecules change to the opposite chirality and thus would reduce
the achieved enantiomeric excess. For some chiral molecules, the effect of tunneling is sufficient weak~\cite{Z.Phys.43.805} or can be suppressed due to the environments~\cite{PRL.88.123001,PRA.91.022709,PR.75.1450,PR.76.1423,PRL.40.980,
MC1,MC2,PRA.99.062703,JCP.112.8743,PRL.103.023202,PRA.88.032504,PCCP.13.17130}.
In our discussions, we have assumed that the tunneling between degenerate chiral states of
different chiralities are negligible~\cite{PRR.2.033064}. Further, the couplings among electronic, vibrational, and rotational degrees of freedom will introduce effective couplings among our work states and other states. Then, the molecules will leak out of the working model, and the few-level model may become insufficient to describe the dynamics of the chiral molecules.
Here, we have assumed that these couplings are negligible in the framework of Born-Oppenheimer approximation~\cite{JCP.151.014302,PRR.2.033064}. These two assumptions will play as main limitations putting on target molecules by using our method of enantio-conversion.}

%Therefore, our optical pumping method
%as well as other enantio-conversion methods~\cite{PRL.90.033001,PRL.84.1669,PRA.65.015401,JCP.115.5349,JPB.37.2811,PRR.2.033064,arXiv2} and enantio-specific state transfer methods~\cite{PRL.87.183002,PRA.77.015403,JPB.43.185402,
%PRL.122.173202,PRA.100.043413,PRAp.13.044021,PRA.101.063401,arxiv1,JCP.151.014302,PRA.100.043403} can only
%apply to these kinds of chiral molecules.

Upon these assumptions and approximations, the vibrational sub-levels of $|1_{L}\rangle$ and $|2_{L}\rangle$ can be chosen as the vibrational ground state in one well of the double-well potential, since the their rotational sub-levels are different. Its inversion image gives the vibrational sub-levels of $|1_{R}\rangle$ and $|2_{R}\rangle$.
We note that the vibrational sub-levels of chiral states $|1_{L}\rangle$ and $|2_{L}\rangle$ can also be chosen as the vibrational ground and higher-energy states in one well of the double-well potential, respectively. The vibrational sub-level of the achiral state $|3\rangle$ can be a vibrational excited state near or beyond the barrier of the double-well potential.

{For gas-phase molecules}~\cite{Nature.497.475,PRL.111.023008,PCCP.16.11114,ACI,JCP.142.214201,
JPCL.6.196,JPCL.7.341,PRL.118.123002,Angew.Chem.56.12512}, the propagation directions of the three  electromagnetic fields can not be parallel due to the special requirement in their polarizations~\cite{PRA.98.043403,JCP.151.014302} (e.g. as in our discussions, the polarization directions are mutually vertical to each other). This gives rise to the
phase-mismatching problem~\cite{Nature.497.475,PRL.111.023008,PCCP.16.11114,ACI,JCP.142.214201,
JPCL.6.196,JPCL.7.341,PRL.118.123002,Angew.Chem.56.12512}. However, when the typical length of the molecule-field interaction volume is much smaller than the largest wavelength among those of the three electromagnetic fields, the molecules are approximately phase-matched. Therefore, throughout our discussions we assume that the molecules are phase matched and described by the same Hamiltonian.

\section{Chiral-state-selective excitations}\label{CSESEC}
For chiral molecules, we will utilize the difference between the overall phases $\phi_{L}$ and $\phi_{R}$ to
establish the required chiral-state-selective excitations. Although the relaxations are indispensable and important in optical
pumping, in order to highlight the physical mechanism of our
chiral-state-selective excitations, we will not consider the relaxations during evolution in this section.

In the large-detuning region with $|\Delta|\gg\Omega_{32}\sim {\Omega_{21}}\gg\Omega_{31}$, we use
the Fr\"{o}hlich-Nakajima transformation~\cite{FST1,FST2} of $\exp{(\hat{S})}$ for the Hamiltonian~(\ref{5HT}).
The anti-Hermitian operator is
\begin{align}
\hat{S}=\frac{1}{\Delta}\sum_{Q}(\Omega_{21}|1_{Q}\rangle\langle 2_{Q}|+\Omega_{32}|3\rangle\langle 2_{Q}|-\mathrm{H.c.}),
\end{align}
which satisfies
$[\hat{H}_{0},\hat{S}]+\hat{H}_{1}=0$. The zero-order, first-order, and second-order Hamiltonian are $\hat{H}_{0}=\sum_{Q}\Delta|2_{Q}\rangle\langle 2_{Q}|$, $\hat{H}_{1}=\sum_{Q}(\Omega_{21}|1_{Q}\rangle\langle 2_{Q}|+\Omega_{32}|2_{Q}\rangle\langle 3|+\mathrm{H.c.})$, and  $\hat{H}_{2}=\sum_{Q}(\Omega_{31}e^{i\phi_{Q}}|1_{Q}\rangle\langle 3|+\mathrm{H.c.})$, respectively.
The transformed Hamiltonian is
\begin{align}\label{THM}
\hat{H}^{\prime}=&\exp{(-\hat{S})}\hat{H}\exp{(\hat{S})}\simeq \hat{H}_{0}+[\hat{H}_1,\hat{S}]/2+\hat{H}_{2}\nonumber\\
=&\sum_{Q}\tilde{\Delta}|2_{Q}\rangle\langle 2_{Q}|
+(\Lambda|2_{L}\rangle\langle 2_{R}|+\mathrm{H.c.})+2\Lambda|3\rangle\langle 3|\nonumber\\
&
+\sum_{Q}[\tilde{\Lambda}|1_{Q}\rangle\langle 1_Q|+
(\tilde{\Omega}_{Q}|1_{Q}\rangle\langle 3|+\mathrm{H.c.})].
\end{align}
Here, we have defined $\Lambda\equiv-\Omega^2_{32}/\Delta$,  $\tilde{\Lambda}\equiv-\Omega^2_{21}/\Delta$, $\tilde{\Delta}\equiv\Delta-\Lambda-\tilde{\Lambda}$, and
\begin{align}\label{PPT}
\tilde{\Omega}_{Q}\equiv\Omega_{31}e^{i\phi_{Q}}-\frac{\Omega_{32}\Omega_{21}}{\Delta}.
\end{align}

{From Eq.~(\ref{THM}), we can see that the evolution of the initial chiral ground states $|1_{Q}\rangle$
will not be affected by the chiral mediate-energy states $|2_Q\rangle$. Thus,
the chiral mediate-energy states $|2_Q\rangle$ can be adiabatically eliminated.
This process gives the following reduced three-level Hamiltonian}
\begin{align}\label{ITF}
\hat{\mathcal{H}}=2\Lambda|3\rangle\langle 3|
&+\sum_{Q}[\tilde{\Lambda}|1_{Q}\rangle\langle 1_Q|+
(\tilde{\Omega}_{Q}|1_{Q}\rangle\langle 3|+\mathrm{H.c.})].
\end{align}

By further tuning the three electromagnetic fields to ensure
\begin{align}\label{CD1}
\phi=0,~~~~\Omega_{31}=\frac{\Omega_{32}\Omega_{21}}{\Delta},
\end{align}
i.e., $\tilde{\Omega}_{L}=0$ and $\tilde{\Omega}_{R}=-2\Omega_{31}$,
{we can arrive the Hamiltonian}
\begin{align}\label{CSER}
\hat{\mathcal{H}}_{\mathrm{eff}}
&=2\Lambda|3\rangle\langle 3|-
2{\Omega}_{31}(|1_{R}\rangle\langle 3|+\mathrm{H.c.})+\sum_{Q}\tilde{\Lambda}|1_{Q}\rangle\langle 1_Q|.
\end{align}
The left-handed ground state $|1_{L}\rangle$
is an eigenstate of the effective Hamiltonian~(\ref{CSER}).
The right-handed ground state $|1_{R}\rangle$
is coupled to $|3\rangle$ in the effective Hamiltonian~(\ref{CSER}).
Therefore, we expect that $|1_{L}\rangle$ will be undisturbed by the electromagnetic fields, and {$|1_{R}\rangle$} will
be excited, i.e., achieving the chiral-state-selective excitations.

{The mediate-energy states $|2_{L}\rangle$ and $|2_{R}\rangle$ are important in establishing the chiral-state-selective excitations. They are introduced to establish the two-photon processes $|1_{Q}\rangle\leftrightarrow|2_{Q}\rangle\leftrightarrow|3\rangle$. The chirality-dependent interferences between the two-photon processes and the one-photon processes $|1_{Q}\rangle\leftrightarrow|3\rangle$ can give rise to the chirality-dependent dynamics for the two enantiomers.
In the large-detuning region, we have adiabatically eliminated the mediate-energy states in the large-detuning region to yield the effective coupling of $\Omega_{32}\Omega_{21}/\Delta$ [i.e., the second term in Eq.~(\ref{PPT})]. The effective Hamiltonian~(\ref{ITF}) clearly shows the chirality-dependent interferences. By further tuning the electromagnetic fields, we can establish the chiral-state-selective excitations as indicated by Hamiltonian~(\ref{CSER}).}

\begin{figure}[h]
  \centering
  % Requires \usepackage{graphicx}
  \includegraphics[width=0.9\columnwidth]{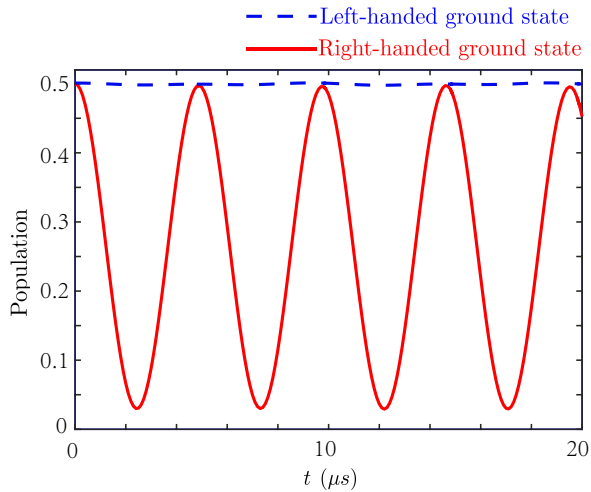}\\
  \caption{Chiral-state-selective excitations: evolution of populations in the left-handed and right-handed ground states. The initial state is ${\rho}(t=0)=\sum_{Q}|1_{Q}\rangle\langle 1_{Q}|/2$. The parameters are $\phi=0$, $\Delta=2\pi\times 20$\,MHz, $\Omega_{21}=2\pi\times1$\,MHz, $\Omega_{32}=2\pi\times1$\,MHz and $\Omega_{31}=2\pi\times0.05$\,MHz. %(b) $\Omega_{32}=2\pi\times1/\sqrt{2}$\,MHz and $\Omega_{31}=2\pi\times0.05/\sqrt{2}$\,MHz.
  }\label{FigP}
\end{figure}
In Fig.~\ref{FigP}, we demonstrate the expected chiral-state-selective excitations
by numerically solving $\dot{{\rho}}=-i[\hat{H},\rho]$ {for an initial racemic mixture with
each molecule described by ${\rho}(t=0)=\sum_{Q}|1_{Q}\rangle\langle 1_{Q}|/2$.}
We take the typical experimental available parameters~\cite{JCP.151.014302,PRR.2.033064}:
$\Delta=2\pi\times 20$\,MHz, $\Omega_{32}=\Omega_{21}=2\pi\times1$\,MHz, $\Omega_{31}=2\pi\times0.05$\,MHz, and $\phi=0$ (more discussions about these parameters are shown in Appendix~\ref{PC}). The numerical results clearly show that
only the right-handed ground state $|1_{R}\rangle$ is excited and the left-handed ground state
$|1_{L}\rangle$ approximately remains unchanged.

%\textcolor{blue}{In Fig.~\ref{FigP}\,(b), we show that our
%well-designed five-level double-$\Delta$ model
%can be used to achieve highly efficient enantio-specific state transfer without the consideration of decoherences by further
%tuning the parameters to ensure $2\Lambda=\tilde{\Lambda}$. For this purpose, the pulse-duration should be precisely controlled as discussed in Ref.~\cite{PRA.100.043403}. Specifically, the pulse should be end when the right-handed ground state experiences
%half-integer periods of its corresponding Rabi oscillation.
%By applying additional well-designed and precisely controlled operations following Refs.~\cite{arXiv2,PRR.2.033064}, we can also achieve highly efficient enantio-conversion.
%In the next section, we will show that with the consideration of decoherences and without
%the requirement of precise control of the pulse-duration, the highly efficient enantio-conversion
%can be achieved via optical pumping. This offers our method advantages in achieving highly efficient inner-state enantio-purification over other theoretical methods~\cite{PRL.90.033001,PRL.84.1669,PRA.65.015401,JCP.115.5349,JPB.37.2811,PRR.2.033064,arXiv2,
%PRL.87.183002,PRA.77.015403,JPB.43.185402,
%PRL.122.173202,PRA.100.043413,PRAp.13.044021,PRA.101.063401,arxiv1,JCP.151.014302,PRA.100.043403}. }

We note that when the {overall phase} $\phi=0$ changes to be $\phi=\pi$, the dynamics of the two chiral ground states exchange to each other~\cite{PRA.100.043403,PRA.100.033411,PRR.2.033064}.
Then, {with the same process as above}, the right-handed ground state $|1_{R}\rangle$ will be undisturbed by the  electromagnetic fields and the left-handed ground state $|1_{L}\rangle$ will
be excited. For convenience, we will only focus on the cases with $\phi=0$ in the following discussions.

\section{enantio-conversion via optical pumping}\label{OP}

Now, we have established the chiral-state-selective excitations without the
consideration of the relaxations. The relaxations are ineluctable in the realistic cases~\cite{PRL.118.123002,Angew.Chem.56.12512,Nature.497.475,PRL.111.023008,PCCP.16.11114,
ACI,JCP.142.214201,JPCL.6.196,JPCL.7.341}, and
important for the realization of our method. In the following, {we will show that when the electromagnetic fields are well designed as in
Sec.~\ref{CSESEC}, the combining effect of the relaxations and the chiral-state-selective excitations can give rise to the enantio-conversion
via optical pumping.}
Now the evolution of {the} density operator ${\rho}$ obeys the master equation
\begin{align}\label{MSE}
\frac{d {\rho}}{d t}=-i[\hat{H},{\rho}]+\mathcal{L}{\rho}.
\end{align}
%\textcolor{red}{2020.7.23}%记得比较和其他方案，确认耗散写对了没有
The effect of decoherences are described by the super-operator $\mathcal{L}{\rho}$. It can be divided into $\mathcal{L}{\rho}=
\mathcal{L}_{\mathrm{rl}}{\rho}+\mathcal{L}_{\mathrm{dp}}{\rho}$. The term of pure population relaxation $\mathcal{L}_{\mathrm{rl}}{\rho}$ reads
\begin{align}\label{REL}
\mathcal{L}_{\mathrm{rl}}{\rho}=&
\sum_{Q}[\gamma_{21}
(\hat{\sigma}_{1_Q2_{Q}}{\rho}\hat{\sigma}_{2_Q1_{Q}}
-\hat{\sigma}_{2_Q1_{Q}}\hat{\sigma}_{1_Q2_{Q}}{\rho})
\nonumber\\
&+\sum^{2}_{n=1}\gamma_{3n}(\hat{\sigma}_{n_{Q}3}{\rho}\hat{\sigma}_{3n_Q}
-\hat{\sigma}_{n_{Q}3}\hat{\sigma}_{3n_{Q}}{\rho})]+\mathrm{H.c.}
\end{align}
with chirality-independent~\cite{JPB.37.2811} relaxation rates $\gamma_{3n}$ and $\gamma_{21}$.
We have defined that $\hat{\sigma}_{pq}\equiv|p\rangle\langle q|$ ($p,q=1_{Q},2_{Q},3$).
We have also considered the effect of pure dephasings, which is also ineluctable in the realistic cases~\cite{Nature.497.475,PRL.111.023008,PCCP.16.11114,ACI,JCP.142.214201,
JPCL.6.196,JPCL.7.341,PRL.118.123002,Angew.Chem.56.12512}. The term of pure dephasing reads~\cite{PD}
\begin{align}\label{DEP}
\mathcal{L}_{\mathrm{dp}}{\rho}=-\tilde{\gamma}\sum_{p,q}\hat{\sigma}_{pp}\rho\hat{\sigma}_{qq},
~(p\ne q).
\end{align}
It only causes the decrease of off-diagonal terms of ${\rho}$.
We have assumed that the pure dephasing rate $\tilde{\gamma}$ is state-independent for the
sake of simplicity~\cite{MP}.

\begin{figure}[h]
  \centering
  % Requires \usepackage{graphicx}
  \includegraphics[width=0.9\columnwidth]{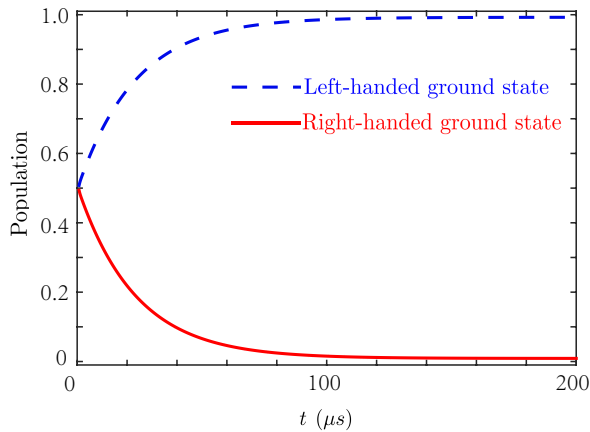}\\
  \caption{Optical pumping method of enantio-conversion: evolution of populations in the left-handed  and right-handed ground states The initial state is ${\rho}(t=0)=\sum_{Q}|1_{Q}\rangle\langle 1_{Q}|/2$. We choose the decoherence rates as $\gamma_{31}=\gamma_{32}=2\pi\times0.1$\,MHz and $\gamma_{21}=\tilde{\gamma}=2\pi\times1$\,MHz. The parameters are $\phi=0$, $\Delta=2\pi\times 20$\,MHz, $\Omega_{21}=2\pi\times1$\,MHz,  $\Omega_{32}=2\pi\times1$\,MHz, and $\Omega_{31}=2\pi\times0.05$\,MHz. %(b) $\Omega_{32}=2\pi\times1/\sqrt{2}$\,MHz and $\Omega_{31}=2\pi\times0.05/\sqrt{2}$\,MHz.
  }\label{Fig2}
\end{figure}

Specifically, we choose the typical decoherence rates in gas-phase experiments~\cite{Nature.497.475,PRL.111.023008,PCCP.16.11114,ACI,JCP.142.214201,
JPCL.6.196,JPCL.7.341,PRL.118.123002,Angew.Chem.56.12512}: $\gamma_{31}=\gamma_{32}=2\pi\times0.1$\,MHz and $\gamma_{21}=\tilde{\gamma}=2\pi\times1$\,MHz (more discussions are shown in Appendix~\ref{PC}). The results by numerically solving master equations~(\ref{MSE}) are shown in Fig.~\ref{Fig2}. They clearly show that the dynamics of the system {are} intensely changed by
the decoherences. The combining effect of the chiral-state-selective excitations and the decoherences results in the conversion of molecules from $|1_{L}\rangle$ to $|1_{R}\rangle$. We are interested in the
enantiomeric excess in the chiral ground states defined as
\begin{align}
\varepsilon\equiv\left|\frac{\langle 1_{L}|{\rho}|1_{L}\rangle-\langle 1_{R}|{\rho}|1_{R}\rangle}{\langle 1_{L}|{\rho}|1_{L}\rangle+\langle 1_{R}|{\rho}|1_{R}\rangle}\right|.
 \end{align}
The achieved enantiomeric excess ($t\ge 140$\,$\mu s$) in Fig.~\ref{Fig2} is $\varepsilon=99.2\%$, i.e.,
highly efficient enantio-conversion {is} achieved.

{\subsection{Role of relaxations and dephasings}\label{IVA}}
Now, we have shown the possibility of highly efficient enantio-conversions by optical pumping.
When the chiral-state-selective excitations and the relaxations act simultaneously on the chiral mixture,
the system gradually reaches its steady state and finally most molecules are transferred to that of the same chirality.
Here, we would like to discuss how the enantio-conversion depends on relaxation and dephasing rates.
For this purpose, we explore the steady states of the system related to the master equation~(\ref{MSE}).
Specifically, we solve the equations $-i[\hat{H},{\rho}]+\mathcal{L}{\rho}=0$.

\begin{figure}
  \centering
  % Requires \usepackage{graphicx}
  \includegraphics[width=0.85\columnwidth]{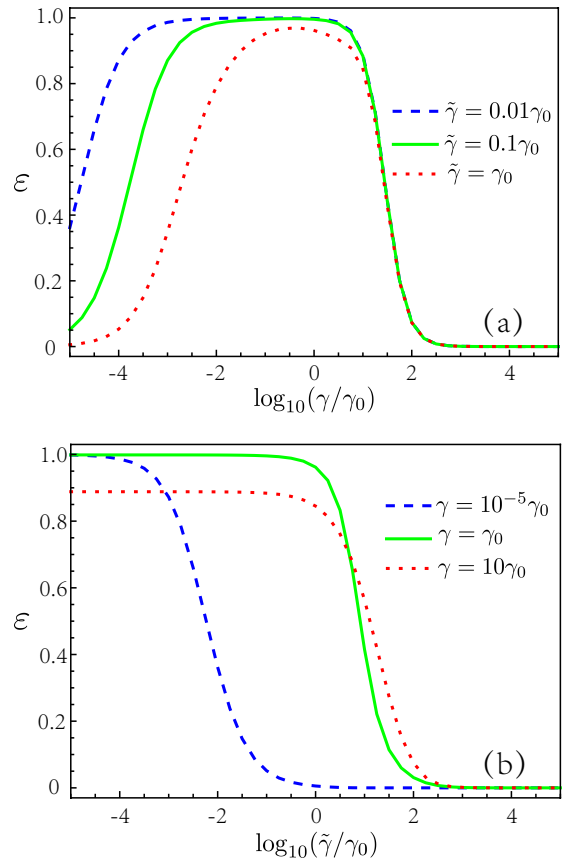}\\
  \caption{Achieved enantiomeric excess $\varepsilon$ in the steady states. $\gamma_0=2\pi\times1$\,MHz.
  The parameters are $\phi=0$, $\Delta=2\pi\times 20$\,MHz, $\Omega_{21}=2\pi\times1$\,MHz,  $\Omega_{32}=2\pi\times1$\,MHz and $\Omega_{31}=2\pi\times0.05$\,MHz.
  (a) $\varepsilon$ as a function of $\gamma~(=\gamma_{31}=\gamma_{32}=\gamma_{21})$ for different dephasing rates $\tilde{\gamma}$. (b) $\varepsilon$ as a function of $\tilde{\gamma}$ for different relaxation rates $\gamma~(=\gamma_{31}=\gamma_{32}=\gamma_{21})$. }\label{FigR}
\end{figure}

For simplicity, we choose $\gamma_{31}=\gamma_{32}=\gamma_{21}=\gamma$.
In Fig.~\ref{FigR}\,(a), we show how the achieved enantiomeric excess $\varepsilon$ (in the steady state) varies with the relaxation rate $\gamma$ for different dephasing rates $\tilde{\gamma}$. {Specifically, we would like to take the case of $\tilde{\gamma}=0.01\gamma_0$ (the blue dashed line) as an example. When $\gamma$ is very small (e.g. $\gamma<10^{-3}\gamma_0$), the achieved enantiomeric excess $\varepsilon$ is not high. In that region, the achieved enantiomeric excess $\varepsilon$ increases with $\gamma$.
In the large relaxation region (e.g. $\gamma>10\gamma_0$),
the achieved enantiomeric excess will decrease with $\gamma$. Then, the highly efficient enantio-conversions can only be achieved in a mediated region of $\gamma$. }

{These phenomena indicate that the relaxations play a complicate role in optical pumping for enantio-conversion. The relaxations make the achiral excited molecules, which was transferred from the undesired chiral ground state due to the chiral-state-selective excitations, relax to the desired chiral ground state. In this sense, the relaxations are indispensable. However, the relaxations should not be very strong, since they may destroy the chiral-state-selective excitations.}

%Generally speaking, the relaxations are indispensable in optical pumping for enantio-conversion, since
%they transfer the achiral excited state to the desired
%chiral ground state. Thus, we expect that the enantiomeric excess in the steady state will not
%high when the relaxation rates are very small.

Further, we compare the results for different dephasing rates in Fig.~\ref{FigR}\,(a).
For the case of small dephasing rate $\tilde{\gamma}=0.01\gamma_0$ (the blue dashed line), highly efficient enantio-conversions (e.g. $\varepsilon\ge 99\%$) can be achieved in a wide region. Here, we have used $\gamma_0=2\pi\times1$\,MHz.
For the case of larger dephasing rate $\tilde{\gamma}=0.1\gamma_0$ (the green solid line), highly efficient enantio-conversions can be achieved in a narrower region. By further increasing the dephasing rate to $\tilde{\gamma}=\gamma_0$ (the blue dashed line), highly efficient enantio-conversions cannot be achieved.

In order to explore the role of the dephasings, we show how the achieved enantiomeric excess $\varepsilon$ varies with $\tilde{\gamma}$ for different relaxation rates ${\gamma}~(=\gamma_{31}=\gamma_{32}=\gamma_{21})$ in Fig.~\ref{FigR}\,(b). It is shown that the achieved enantiomeric excess will
decrease with the increase of the dephasing rates $\tilde{\gamma}$.

{\subsection{Highly efficient enantio-conversion in strong relaxation region}}
\begin{figure}[h]
  \centering
  % Requires \usepackage{graphicx}
  \includegraphics[width=0.9\columnwidth]{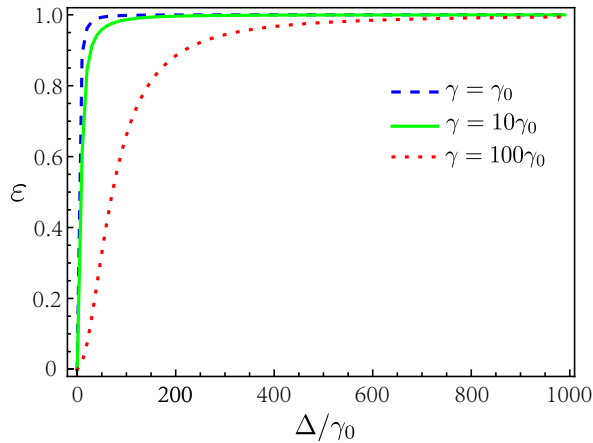}\\
  \caption{Achieved enantiomeric excess in the steady states as a function of $\Delta$ for different $\gamma~(=\gamma_{21}=\gamma_{32}=\gamma_{31})$. Other parameters are chosen as
  $\tilde{\gamma}=2\pi\times1$\,MHz, $\Omega_{21}=2\pi\times1$\,MHz, and $\Omega_{32}=2\pi\times1$\,MHz.
  The coupling strength $\Omega_{32}$ is adjusted correspondingly to ensure $\Omega_{31}=\Omega_{32}\Omega_{21}/\Delta$. We use $\gamma_0=2\pi\times1$\,MHz.}\label{FigS}
\end{figure}

{We have shown that the relaxations are indispensable for enantio-conversion
via optical pumping, but should be not too strong to destroy the chiral-state-selective excitations.
Here, we would like to discuss the possibility of achieving highly efficient enantio-conversion in the
strong relaxation region.}

For this purpose, we increase the detuning $\Delta$ and simultaneously decrease the coupling strength $\Omega_{31}$ to ensure $\Omega_{31}=\Omega_{32}\Omega_{21}/\Delta$.
In Fig.~\ref{FigS}, we show the achieved enantiomeric excess as a function of $\Delta$ for different $\gamma~(=\gamma_{21}=\gamma_{32}=\gamma_{31})$. Other parameters are chosen as
$\tilde{\gamma}=2\pi\times1$\,MHz, $\phi=0$, $\Omega_{21}=2\pi\times1$\,MHz and $\Omega_{32}=2\pi\times1$\,MHz.
The numerical results show that highly efficient enantio-conversion can be achieved in the strong relaxation region. \textcolor{blue}{These results can be understood as follows.
In the large-detuning region with $|\Delta|\gg\Omega_{32}\sim {\Omega_{21}}\gg\Omega_{31}$, the three-level Hamiltonian~(\ref{ITF}) can be used to approximately replace the system's original Hamiltonian~(\ref{5HT}). By further adjusting the fields, we achieve the chiral-state-selective excitation as shown in the system's approximate Hamiltonian~(\ref{CSER}). The larger the detuning $\Delta$ is (or the smaller coupling strengths are), the more robust the approximation is. When a system's Hamiltonian is exactly described by Eq.~(\ref{CSER}), the undesired chiral ground state can be perfectly transferred to the desired chiral ground state by optical pumping in the steady state. In this sense, the larger detuning $\Delta$ is (or the smaller coupling strengths are), the higher achieved enantiomeric excess is.}

{\subsection{Effect of relaxations out of the five-level model }}
\begin{figure}[h]
  \centering
  % Requires \usepackage{graphicx}
  \includegraphics[width=0.9\columnwidth]{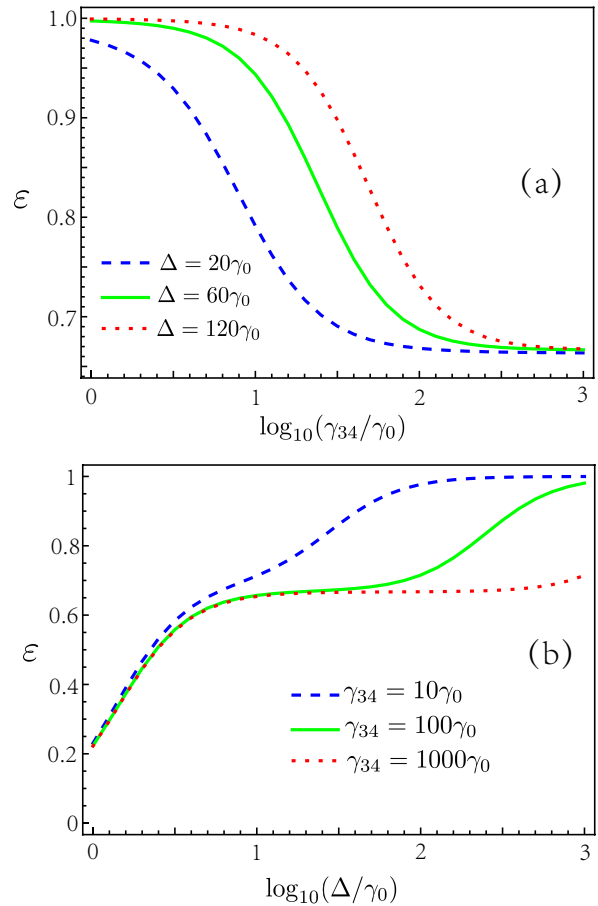}\\
  \caption{Achieved enantiomeric excess in steady states as a function of (a) $\gamma_{34}$ and (b) $\Delta$. The coupling coupling strengths are simultaneously adjusted to ensure $\Omega_{31}=\Omega_{32}\Omega_{21}/\Delta$ in the appearance of relaxation out of the five-level model. Other parameters are $\gamma_{41}=2\pi\times10^{-5}$\,MHz, $\gamma_{31}=\gamma_{32}=2\pi\times0.1$\,MHz, $\gamma_{21}=\tilde{\gamma}=2\pi\times1$\,MHz, $\phi=0$, and $\Omega_{21}=\Omega_{32}=2\pi\times1$\,MHz. We use $\gamma_0=2\pi\times1$\,MHz.}\label{Fig7l}
\end{figure}
{So far}, we have assumed that the excited molecules will not relax out of the five-level model {in our calculations and discussions}. In the realistic case, the molecules in the achiral excited state can relax to {the lower-energy} chiral states out of the five-level model~\cite{PRL.116.063006}. The relaxation rates from these chiral states to their corresponding chiral ground states may be very small~\cite{PRL.116.063006}. Here, we would like to explore the effect of such a process on the optical pumping for enantio-conversion.

To this end, we introduce
additional chiral states $|4_L\rangle$ and $|4_R\rangle$. We describe the relaxation and the dephasing related to them in the same manner as Eq.~(\ref{REL}) and Eq.~(\ref{DEP}), respectively.
For relaxations from $|4_Q\rangle$ to $|1_{Q}\rangle$ ($Q=L,R$), we choose a very small relaxation rate
$\gamma_{41}=2\pi\times10^{-5}$\,MHz. The relaxation rates from the achiral excited state $|3\rangle$ to $|4_{Q}\rangle$ is describe by $\gamma_{34}$.

\textcolor{blue}{
In Fig.~\ref{Fig7l}(a), we show the achieved enantiomeric excess in the steady states as a function of $\gamma_{34}$ for different detuning $\Delta$. We find that the
achieved enantiomeric excess $\varepsilon$ deceases with the increase of the relaxation rate $\gamma_{34}$
and approaches to a constant value ($\sim60\%$). Comparing the results for different detunings, we find that
the achieved enantiomeric excess can be improved by increasing the detuning $\Delta$. In Fig.~\ref{Fig7l}(b), we show how the increase of the detuning $\Delta$ can increase the achieved enantiomeric excess. In Fig.~\ref{Fig7l}, the coupling strengths are simultaneously adjusted to ensure $\Omega_{31}=\Omega_{32}\Omega_{21}/\Delta$. Other parameters are $\gamma_{31}=\gamma_{32}=2\pi\times0.1$\,MHz, $\gamma_{21}=2\pi\times1$\,MHz, $\phi=0$, and $\Omega_{21}=\Omega_{32}=\tilde{\gamma}=2\pi\times1$\,MHz.}

\section{Summary}

We have used the five-level double-$\Delta$ model of chiral molecules to
demonstrate that the highly efficient enantio-conversion can be achieved when
the system reaches its steady state via optical pumping. The chiral-state-selective
excitations
can be generated by well designing the relative phases and intensities among the
three electromagnetic fields with the help of the adiabatical elimination technology.
{The relaxations among our working states are indispensable in our method.
When the relaxations and the chiral-state-selective excitations simultaneously act on
the chiral mixture, the highly efficient enantio-conversion can be achieved.
However, the relaxations should not be too strong to destroy the chiral-state-selective excitations.
We also find that the efficiency of the enantio-conversion will decrease with the increase of the dephasing rates.
In addition, the relaxation out of the working states can also reduce the efficiency of the enantio-conversion via optical pumping. Fortunately, our numerical results have shown that these negative effects can be weakened by
increasing the detunings $\Delta$ and simultaneously reducing the coupling strength $\Omega_{31}$ to ensure $\Omega_{31}=\Omega_{32}\Omega_{21}/\Delta$. In addition, in the whole process of optical pumping, the
relative phases and strengths of the three electromagnetic fields should be well controlled.}

{One key point of our optical pumping method for enantio-conversions is
using the left-right symmetry-breaking in $\Delta$-type (sub-)structures to
establish the chiral-state-selective excitations. The other key point is
the relaxations from the achiral excited state to the chiral ground states.
In our five-level double-$\Delta$ model with one achiral excited state, we have appropriately chosen the applied electromagnetic fields
in the large-detuning condition such that the original model can be reduced to a three-level model. In the reduced three-level model with one achiral excited state, the achiral excited state is coupled with the two chiral ground states on-resonance. By further adjusting the applied electromagnetic fields, the coupling strength corresponding to one chiral ground state can be zero and simultaneously that corresponding to the other one is nonzero, i.e,
achieving the chiral-state-selective excitations. When the chiral-state-selective excitations and the relaxations act simultaneously on the chiral mixture, the enantio-conversion via optical pumping can be achieved.
In this sense, the four-level double-$\Delta$ model~\cite{PRL.84.1669,PRA.65.015401,JCP.115.5349,JPB.37.2811,PRR.2.033064,arXiv2} can also be used to realise enantio-conversions via optical pumping, since it have chirality-dependent $\Delta$-type (sub-)structures and achiral excited states. In Ref.~\cite{PRR.2.033064}, it was shown that
the four-level double-$\Delta$ model~\cite{PRL.84.1669,PRA.65.015401,JCP.115.5349,JPB.37.2811,PRR.2.033064,arXiv2}
can be simplified to two uncoupled two-level subsystems by appropriately choosing the applied fields. The two uncoupled two-level subsystems have the same coupling strengths but different detunings~\cite{PRR.2.033064}. By further adjusting the applied fields, one subsystem will be on-resonance and the other one will be in large-detuning limit. Then, one chiral ground state will be excited and the other will be undisturbed, i.e., achieving the chiral-state-selective excitations. Similar as the case of the five-level model, the chiral-state-selective excitations and the relaxations act simultaneously on the chiral mixture and can eventually evoke the enantio-conversion via optical pumping in the four-level double-$\Delta$ model~\cite{PRL.84.1669,PRA.65.015401,JCP.115.5349,JPB.37.2811,PRR.2.033064,arXiv2}.}\\

\section*{Acknowledgement}
This work was supported by the National Key R\&D Program of China grant (2016YFA0301200), the Natural Science Foundation of China (under Grants No.~11774024, No.~12074030, No.~U1930402 and No.~11947206), and the Science Challenge Project (under Grant No.~TZ2018003).

\appendix*
%\appendices
\section{\textcolor{blue}{Typical parameters for gas-phase chiral molecules}}\label{PC}
Here, we give typical parameters for gas-phase chiral molecules according to
the gas-phase experiments of the enantio-discrimination~\cite{Nature.497.475,PRL.111.023008,PCCP.16.11114,ACI,JCP.142.214201,
JPCL.6.196,JPCL.7.341} and the enantio-specific state transfer~\cite{PRL.118.123002,Angew.Chem.56.12512}.
In our method, the transitions $|3\rangle\leftrightarrow|1_{Q}\rangle$ and
$|3\rangle\leftrightarrow|2_{Q}\rangle$ are ro-vibrational transitions.
The transitions $|2_{Q}\rangle\leftrightarrow|1_{Q}\rangle$ can be purely rotational transitions or ro-vibrational transitions.
Typical experimentally available coupling strengths for rotational transitions of
chiral molecules are about $2\pi\times10$\,MHz or less~\cite{Nature.497.475,PRL.111.023008,PCCP.16.11114,ACI,JCP.142.214201,
JPCL.6.196,JPCL.7.341,PRL.118.123002,Angew.Chem.56.12512}. Usually, the
electric-dipole moments of ro-vibrational transitions are much smaller
than that of purely rotational transitions. Since the intensity of an infrared
electromagnetic field can {be easily made} much larger than that of microwave ones,
the coupling strengths of ro-vibrational transitions can have the same
magnitude as that of purely rotational transitions~\cite{JCP.151.014302}.

{In our calculations, the coupling strengths ($\le 2\pi\times 1$\,MHz) are chosen to be  much smaller than
the experimental available ones ($2\pi\times10$\,MHz).} The reason is {as follows}.
When the electromagnetic fields are applied to manipulate the
chiral molecules, the sample can also be heated. This will
prevent the realization of our method as well as the enantio-discrimination~\cite{Nature.497.475,PRL.111.023008,PCCP.16.11114,ACI,JCP.142.214201,
JPCL.6.196,JPCL.7.341} and the enantio-specific state transfer~\cite{PRL.118.123002,Angew.Chem.56.12512}.
The detunings in our calculations are $\Delta_{21}=-\Delta_{32}=\Delta=2\pi\times20$\,MHz and $\Delta_{13}=0$. The detunings and coupling strengths are much smaller than the typical ro-vibrational and rotational transition frequencies of chiral molecules. Then, the rotating-wave approximation can be used in our model~\cite{PRR.2.033064} and the effect of other selection-rule-allowed transitions of chiral molecules on
our model is negligible~\cite{PRR.2.033064}.

The parameters of decoherences are chosen {as follows}.
The decoherence mechanisms of chiral molecules in gas-phase are
spontaneous emission and collision. For rotational transitions, the
collision dominates and gives the typical rotational relaxation and dephasing rates
of about $2\pi\times1$\,MHz~\cite{Nature.497.475,PRL.111.023008,PCCP.16.11114,ACI,JCP.142.214201,
JPCL.6.196,JPCL.7.341,PRL.118.123002,Angew.Chem.56.12512}.
Since the dephasing varies modestly with molecule, vibrational state, and rotational state~\cite{MP},
we have chosen the state-independent dephasing rate $\tilde{\gamma}=2\pi\times1$\,MHz.
Due to the large transition frequencies, the typical ro-vibrational relaxation rates due to collision are much smaller than the rotational relaxation rates~\cite{MP}.
For ro-vibrational transitions, the relaxations due to spontaneous emission
becomes a possible relaxation mechanism, in addition to collisions with other molecules~\cite{JCP.151.014302}. Therefore, we have chosen the relaxation rate for ro-vibrational
transitions as about $2\pi\times0.1$\,MHz. Specifically, we have chosen $\gamma_{31}=\gamma_{32}=2\pi\times0.1$\,MHz corresponding to the two ro-vibrational transitions.
Since the transitions $|1_{Q}\rangle\leftrightarrow|2_{Q}\rangle$ can be
{purely rotational}, we have chosen $\gamma_{21}=2\pi\times1$\,MHz.

{}

\end{document}